\documentclass{actaps}
\usepackage{fleqn,times,graphics}

\begin{document} 

\headings{1}{8}                             
\def\authorlist{C. Henkel, M. Wilkens}      
\def\shorttitle{Heating of trapped particles close to surfaces}

\title{\uppercase{Heating of trapped particles close to surfaces --- Blackbody
and beyond}}

\author{C. Henkel\email{Carsten.Henkel@quantum.physik.uni-potsdam.de}, 
M. Wilkens}
{
Institute of Physics, University of Potsdam,
Am Neuen Palais 10, 14469 Potsdam, Germany
}

\datumy{10 May 2000}{15 May 2000}       

\abstract{%
We discuss heating and decoherence in traps for ions and neutral particles
close to metallic surfaces. We focus on simple trap geometries and 
compute noise spectra of thermally excited electromagnetic fields.
If the trap is located in the near field of the substrate, 
the field fluctuations are largely increased
compared to the level of the blackbody field, leading to much shorter
coherence and life times of the trapped atoms. The corresponding time constants
are computed for ion traps and magnetic traps. Analytical estimates for the
size dependence of the noise spectrum are given. We finally discuss prospects
for the coherent transport of matter waves in integrated surface waveguides.
}

\medskip


\section{Introduction}

In the field of particle cooling and trapping, a strong trend towards
miniaturisation and integration has emerged in the last few years. 
Small particle traps might form the building blocks of future quantum
computers. Integrated atom optical circuits might distribute 
coherently matter waves for interferometric or nanolithographic
applications. Major issues in this field are heating and decoherence
that have to be controlled in order to maintain the coherence properties
as long as possible. Heating is an intriguing concern because of 
the dramatic temperature difference between the
trapped particles and the macroscopic objects that form a typical
miniature trap. In fact, recent experiments with small 
ion traps have revealed that the life time of the ion's vibrational
ground state gets shorter in smaller traps, making it difficult to down 
scale the trap geometry to micrometre size \cite{Wineland98,Monroe00}. 
In neutral particle traps, sizes in the micrometre range have already
been achieved experimentally \cite{Ovchinnikov97b,Mlynek98b,Haensch98,Hinds98},
but to our knowledge, life times are difficult to measure and are rather 
limited by background pressure and inelastic few-body collisions.

In this contribution, we consider a simplified geometry for particle
traps and compute heating and decoherence rates. The work is divided
in two parts. We start with the heating of the vibrational motion of
an ion in a tightly confining potential. Using perturbation
theory, the heating rate is linked to the cross correlation
spectrum of the electric field at the trap center. This noise spectrum
is evaluated asymptotically, taking into account that the trap distance 
is small  compared to the photon wave length associated with the ion's 
vibration frequency. As a by-product, we also get the noise spectrum
for the magnetic field. This spectrum determines the life time of a
neutral particle in a magnetic trap, since the fluctuating field 
induces spin transitions to a non-trapped state,
kicking the particle out of the trap. In the second part, we focus
on the quasi-free motion of a particle in a wave guide (linear or
planar). The particle scatters from thermal field fluctuations
and thus loses its spatial coherence. A transport theory is formulated
and analytically solved in the limit of a broad-band fluctuation
spectrum.

\section{Heating of a trapped ion}

\subsection{Heating rate}

Let us focus on a single degree of freedom of the ion's motion 
and assume a parabolic confining potential. The Hamiltonian is 
then simply given by
\be
H_{\rm trap} = \hbar\Omega \left( b^\dagger b + {\textstyle \frac12} \right)
\label{eq:1}
\ee
with $\Omega$ the ion's vibration frequency (typically in the MHz range).
If the ion is perturbed by a time-dependent force, this is described 
by the potential \cite{Lamoreaux97,James98}
\be
V(t) = - {\bf x} \cdot {\bf F}( {\bf r}, t )
=
- a \hat {\bf n} \cdot {\bf F}( {\bf r}, t )
\left( b^\dagger + b \right)
.
\label{eq:2}
\ee
We assume that the force varies on a spatial scale much larger than
the size $a = (\hbar/(M \Omega))^{1/2}$ of the trap ground state
and evaluate it at the trap center ${\bf r}$.
${\bf x} = x \hat{\bf n}$ is the ion's displacement from the center.

Using standard second-order perturbation theory, one may easily derive
a master equation for the reduced density matrix of the ion, that
describes its dynamics when the fluctuations of the force field
${\bf F}( {\bf r}, t)$ are traced over \cite{Lamoreaux97,James98,Wineland98,%
Henkel99c}. This master equation allows
to derive the equation of motion for the population of the lowest
trap levels, as well as for the average creation operator and 
the average excitation number (in the absence of cooling processes)
\begin{eqnarray}
\dot \rho_{00} & = & - \gamma_- \rho_{00} + \gamma_+ \rho_{11}
\label{eq:3}
\\
\langle \dot n \rangle & = & 
- (\gamma_+ - \gamma_-) \langle n \rangle 
+ \gamma_-
\label{eq:4}
\\
\langle \dot b \rangle & = & 
- i \Omega \langle b \rangle 
- \frac12 ( \gamma_+ - \gamma_- ) \langle b \rangle 
\label{eq:5}
\end{eqnarray}
The rates $\gamma_{\pm}$ in these formulas are related to the
cross spectral density of the force fluctations as follows
\begin{eqnarray}
\gamma_{\pm} & = & \gamma( {\bf r}; \pm\Omega )
\nonumber\\
& = &
\frac{ a^2 }{ \hbar^2 }
\sum_{i,j} \hat n_i \hat n_j S_F^{ij}( {\bf r}; \pm \Omega )
\label{eq:6}
\end{eqnarray}
with
\be
S_F^{ij}( {\bf r}; \omega ) =
\int_{-\infty}^{+\infty} \! d\tau \,
\langle F_i( {\bf r}, t + \tau ) 
F_j( {\bf r}, t ) \rangle
e^{i \omega \tau }
\label{eq:def-spectral-density}
\ee
According to~(\ref{eq:3}), the heating rate from the trap ground state is 
thus given by $\Gamma_{0\to 1}( {\bf r}) = \gamma_-$. We are thus left
with the calculation of the spectral density~(\ref{eq:def-spectral-density}) 
for the perturbing force.

\subsection{Electric field noise spectrum}

The trapped ion being charged, it is perturbed by fluctuating electric 
fields, ${\bf F} = q {\bf E}$. The Planck blackbody spectrum then gives
the following spectral density
\be
S_F^{ij}( {\bf r}; \omega ) =
q^2 S_E( \omega ) \delta^{ij} 
=
\frac{ q^2 \, \hbar \omega^3 \, \delta^{ij} }{
3 \pi \varepsilon_0 c^3 ( 1 - {\rm e}^{-\hbar\omega / k_B T } ) }
\label{eq:7}
\ee
where $T$ is temperature.
One must keep in mind, however, that this gives the thermal spectrum only  
in free space, far from the sources. But the ion is trapped at a distance
$z$ from the trap electrodes that is typically much smaller than the
photon wave length $\lambda = 2\pi c / \Omega$ associated with the
vibration frequency. It is thus located in the near field of the
electrodes, and the Planck formula does not cover this case.
This was recognised already in the early days of ion trapping when
`hot' ion clouds (with temperatures much larger than room temperature)
were cooled down by thermalisation with the surrounding electrodes,
the coupling being provided by the absorption of the electric fields
radiated by the moving ions in the lossy metallic environment
\cite{Wineland75}. 
When laser cooling took over to reach temperatures in the $\mu$K range,
voltage fluctuations due to electric losses became a source of heating.
Modeling the ion trap as a lumped circuit with resistance $R(\omega)$,
the standard Nyquist formula for Johnson noise gives an electric
field spectrum \cite{Lamoreaux97,James98,Wineland75}
\be
S_F( {\bf r}; \omega ) 
\approx
\frac{ q^2 \, k_B T R(\omega) }{ z^2 }
\label{eq:Johnson-noise}
\ee
where the high-temperature limit $k_B T \gg \hbar |\omega|$ has been
assumed. For a given trap geometry, it is not easy to determine the 
resistance $R(\omega)$ that enters this formula. Assuming that the
electric currents propagate only in the skin layer of the electrodes,
the NIST group estimated that the spectrum~(\ref{eq:Johnson-noise}) 
actually
gives a heating rate too small to account for the experimental observations.
In addition, there are indications that the scaling law 
$\Gamma_{0\to 1}( {\bf r} ) \propto S_F({\bf r}) \propto 1/z^2$ 
is not followed experimentally (although it is difficult to exclude
other influences when down-scaling the trap geometry) \cite{Monroe00}.

We now outline a microscopic effective model \cite{Henkel99b,Henkel99c,%
Monroe00} that may in principle allow to compute the electric 
noise spectrum for an arbitrary trap geometry.
\begin{figure}
\begin{center}
\resizebox{6.0cm}{!} {\rotatebox{0}{
   \includegraphics{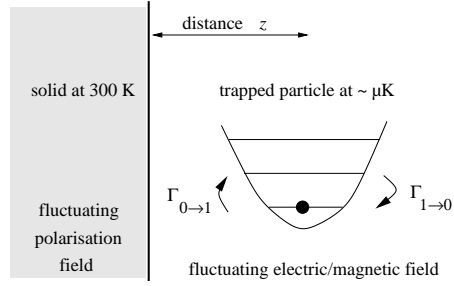}
   }}
\end{center}
\caption{{\bf Fig.\ 1.}
Sketch of the microscopic effective model used to compute the electric
field noise spectrum.}
\end{figure}
We describe the electric properties of the surrounding metal by 
its frequency-dependent complex dielectric function 
$\varepsilon( {\bf r}; \omega)$. The microscopic source of Johnson
noise are fluctuating polarisation fields ${\bf P}( {\bf r}, t)$
that are thermally excited
in the metal. According to the fluctuation dissipation theorem, the
spectral density of this fluctuating polarisation is related to the 
imaginary part of the dielectric function 
\cite{Lifshitz56,Barnett92,Knoell98a}:
\be
S_P^{ij}( {\bf r}', {\bf r}; \omega ) =
\frac{ 2 \hbar \varepsilon_0 \, {\rm Im}\,\varepsilon( {\bf r}; \omega ) }
{ 1 - {\rm e}^{ - \hbar \omega / k_B T } }
\delta^{ij}
 \delta( {\bf r}' - {\bf r} )
\label{eq:pol-field}
\ee
Maxwell's equations now determine the field radiated by this
polarisation. It may be written as an integral over the Green
tensor
\be
E_i( {\bf r}; \omega ) =
\int\!{\rm d}{\bf r}' \sum_{j}
G^{ij}( {\bf r}, {\bf r}'; \omega )
P_j( {\bf r}'; \omega )
\label{eq:field-and-Green}
\ee
where $G^{ij}$ depends on the geometry of the trap electrodes.%
\footnote{Rigorously speaking, the electric field also contains
a contribution due to modes impinging from infinity in the vacuum
space. In the short distance limit $z \ll \lambda$ relevant here,
however, this contribution may be shown to be very small
(C. Henkel, K. Joulain, R. Carminati, J.-J. Greffet, in preparation).}

To proceed with the calculation, we now fix the geometry and
consider a trap located at a distance $z$ above an infinite flat
metallic surface. In this geometry, the Green function 
in~(\ref{eq:field-and-Green}) is explicitly known in spatial 
Fourier space, and it is possible to perform an asymptotic expansion
in the near field limit $z \ll \lambda$. As a result, we obtain the
following interpolation formula \cite{Monroe00,Henkel99c}
\be
S_E^{ij}( z ; \omega ) \approx 
\frac{ \hbar \omega \, \varrho }{
4 \pi ( 1 - {\rm e}^{ - \hbar \omega / k_B T } ) \, z^3 }
\left(
s^{ij} + \delta^{ij} \frac{ z }{ \delta( \omega ) } 
\right)
.
\label{eq:electric-asymptotics}
\ee
As expected, the spectrum only depends on the $distance$ from the surface.
The material properties enter through the specific resistance $\varrho$
and the skin depth $\delta( \omega ) = c \sqrt{ \varepsilon_0 \varrho 
/ \omega }$. The geometry enters through the power law $S_E(z) \propto 1/z^3$
in the `extreme near field' $z \ll \delta$. This behaviour may be
understood from the principle of detailed balance: the heating rate
of the ion is equal to the relaxation rate of an oscillating electric 
dipole, and it is well known that close to a metallic surface, this
rate is dominated by nonradiative transfer and increases 
as $1/z^3$ \cite{Drexhage74,Scheel99b}. 

We may also extract 
from~(\ref{eq:electric-asymptotics}) an `effective resistance':
if the distance is large compared to the skin depth, the near field spectrum
indeed follows a $1/z^2$ power law, as in eq.(\ref{eq:Johnson-noise}).
The comparison yields an effective resistance
$R_{\rm eff}( \omega ) \approx 3 \varrho / ( 4 \pi \delta ( \omega ) )$.
This indicates that the thermal currents actually flow in a skin
layer right below to the metallic surface, in agreement with
the estimates made by the NIST group \cite{Wineland98}.

The ion heating rate obtained from the 
spectrum~(\ref{eq:electric-asymptotics}) is plotted in figure~2.
\begin{figure}
\begin{center}
\resizebox{8.0cm}{!} {\rotatebox{0}{
   \includegraphics{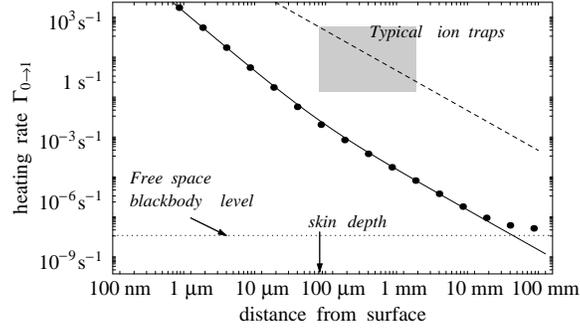}
   }}
\end{center}
\caption{\label{fig:2}%
{\bf Fig.\ 2.}
Heating rate for a trapped ion (mass $40\,$amu, charge $q = e$) trapped
in a harmonic trap (frequency $\Omega/2\pi = 1\,$MHz) above a copper
surface at $300\,$K. Dots: exact evaluation of the electric near field 
fluctations; solid line: asymptotic expansion~(\ref{eq:electric-asymptotics});
dashed line: heating rate from the Johnson noise 
spectrum~(\ref{eq:Johnson-noise}) with $R(\omega) = 1\,\Omega$.}
\end{figure}
We observe that the electric near field fluctuates at a noise level
much larger than the blackbody field (dotted line). It is also apparent
that life times shorter than 1~s are to be expected when the trap size
gets into the micrometre range. The figure also shows that the life times
in current traps are not limited by the near field fluctuations discussed
here, the observed heating rates being much larger. The NIST group proposed
a model including fluctuating electric patch potentials that may explain
this discrepancy, but little is known about the dielectric properties of
these electric surface domains in the relevant frequency range
\cite{Monroe00}. In view of recent theoretical and experimental
investigations \cite{Wineland98,Monroe00,Henkel99c,Henkel99b,Knight98,%
Milburn98}, a detailed understanding of ion heating processes is still 
lacking.

\subsection{Magnetic field noise spectrum}

To conclude this section, we turn to a different type of trap: a static
inhomogeneous magnetic field that traps a paramagnetic atom at local
minima of the magnetic field strength. Such traps are routinely used for
evaporative cooling \cite{Cornell95b}, and miniature versions close
to wires or surfaces have been proposed \cite{Hinds98,Schmiedmayer95} 
and actually realised \cite{Schmiedmayer98b,Reichel99,Anderson99,%
Prentiss00,Hinds00}. Thermally 
fluctuating magnetic fields now may flip the atomic spin, leaving the
atom possibly in an unbound potential (see figure~3).
\begin{figure}
\begin{center}
\resizebox{4.5cm}{!} {\rotatebox{0}{
   \includegraphics{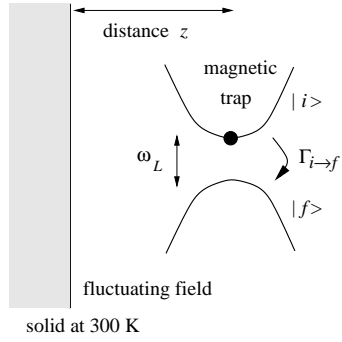}
   }}
\end{center}
\caption{{\bf Fig.\ 3.}
Fluctuating magnetic fields flip the spin of an atom in a magnetic trap
and put it onto a non-trapping potential surface.}
\label{fig:3}
\end{figure}
The spin flip rate is easily obtained from perturbation theory and
relates to the spectral density of the magnetic field fluctuations:
\be
\Gamma_{i\to f}( {\bf r} ) =
\frac{ 1 }{ \hbar^2 }
\sum_{\alpha,\,\beta}
\langle i | \mu_\alpha | f \rangle
\langle f | \mu_\beta | i \rangle
S^{\alpha\beta}_B( {\bf r}; \omega_L )
\label{eq:9}
\ee
where $\omega_L = \mu B_{\rm trap}( {\bf r} ) / \hbar$ is the Larmor
frequency. The magnetic field spectrum may be calculated using the
theory outlined above, and one finds the following asymptotic result
\cite{Henkel99c}:
\be
S^{\alpha\beta}_B( z; \omega ) \approx
\frac{ \hbar \omega \, s^{\alpha\beta} }{
16 \pi \varepsilon_0^2 c^4 \varrho\,
( 1 - {\rm e}^{ - \hbar\omega / k_B T } ) 
\,z}
\left(
1 + \frac{ 2 z^3}{ 3 \delta^3( \omega ) } 
\right)^{-1}
\label{eq:magnetic-spectrum}
\ee
For a spin 1/2 particle, we thus get the trap loss rate shown in
figure~4.
\begin{figure}
\begin{center}
\resizebox{8.0cm}{!} {\rotatebox{0}{
   \includegraphics{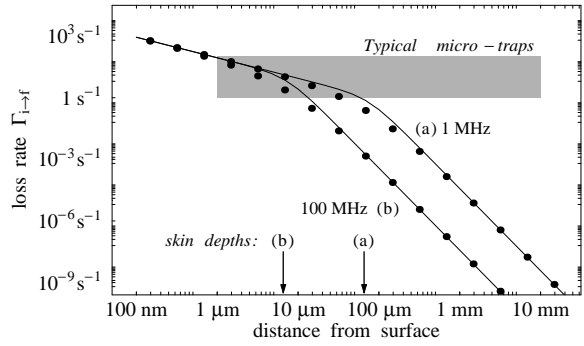}
   }}
\end{center}
\caption{{\bf Fig.\ 4.}
Spin flip rate (loss rate) for a paramagnetic atom 
(magnetic moment $\mu = \mu_B$) in a magnetic trap
above a copper surface at $300\,$K. The Larmor frequency takes the values
$\omega_L/2\pi = 1\,$MHz (curve (a)) and $100\,$MHz (curve (b)).
The rate obtained from the blackbody spectrum is much smaller 
(about $10^{13}\,{\rm s}^{-1}$ for $\omega_L/2\pi = 100\,$MHz).
}
\label{fig:4}
\end{figure}
We note that in micrometre size traps, the life time is limited
to less than a second due to near field fluctuations, which should
be an observable effect in current experiments.

\section{Decoherence in wave guides}

We now turn to the influence of fluctuating near fields on the quasi-free
motion of atoms in linear or planar wave guides  
\cite{Mlynek98b,Hinds98,Schmiedmayer95,Schmiedmayer98b,Hinds00}.
The atomic matter wave is scattered from spatial inhomogeneities of 
the perturbing field, thus changing its momentum. We assume again
a statistical description of the scattering process. 
The typical momentum transfer is thus of the order of $\hbar/\ell$
where $\ell$ is the correlation length of the field. Energy is not
conserved because the field is fluctuating, and the maximum energy
transfer is roughly limited by $k_B T$.
Note that this is typically much larger than the kinetic energies 
involved in cold atomic clouds.

\subsection{Transport equation}

We want to characterise the evolution of the single-particle spatial
density matrix (or coherence function)
\be
\rho( {\bf r}; {\bf s} ) = 
\langle
\psi^*( {\bf r} + {\textstyle\frac12} {\bf s} )
\,
\psi( {\bf r} - {\textstyle\frac12} {\bf s} )
\rangle
\label{eq:10}
\ee
where the average $\langle \ldots \rangle$ is taken over the spatial
and temporal fluctuations of a perturbing potential $V( {\bf r}, t)$.
It is useful to introduce the Wigner transform of the density matrix
\be
W( {\bf r}, {\bf p} ) = \int \!
\frac{ d^3s }{ 2 \pi\hbar } e^{ i {\bf p} \cdot {\bf s} / \hbar }
\rho( {\bf r}; {\bf s} )
\label{eq:11}
\ee
that may be interpreted as a quasi-probability distribution in phase
space.

Using second-order perturbation theory, assuming gaussian statistics
for the perturbing potential and doing a multiple-scale
expansion of the Bethe-Salpeter equation for the coherence function,
we get the following transport equation \cite{Keller96}:
\begin{eqnarray}
&&
\Bigl(
\partial_t + 
\frac1m {\bf p} \cdot \nabla_{\bf r} 
+ 
{\bf F}_{\rm ext} \cdot \nabla_{\bf p}
\Bigr)
W( {\bf r}, {\bf p} )
=
\nonumber\\
&& 
\int\!{\rm d}^Dp' \,
S_V( {\bf p}' - {\bf p}; E_{p'} - E_{p} )
\left[
W( {\bf r}, {\bf p}' )
-
W( {\bf r}, {\bf p} )
\right]
\label{eq:transport-eq}
\end{eqnarray}
where $D = 1,2$ is the dimension of the wave guide and $S( {\bf q}; \Delta E)$
the spectral density of the fluctuating potential
\be
S_V( {\bf q}; \Delta E) =
\frac{ 1 }{ \hbar^2 }
\int\!\frac{ d^D s \, d\tau}{ (2\pi\hbar)^{D} }
\langle
V( {\bf r} + {\bf s}, t + \tau) \,
V( {\bf r}, t ) 
\rangle
\, e^{ - i ( {\bf q}\cdot{\bf s} - \Delta E \tau ) / \hbar}
.
\label{eq:13}
\ee
We have assumed that the potential is statistically stationary in both
space and time. 

The left hand side of the transport equation~(\ref{eq:transport-eq})
describes the ballistic motion of the atom subject to the external 
(deterministic) force ${\bf F}_{\rm ext}$. The right hand side
describes the scattering off the fluctuating potential. 
As a function of the momentum transfer $\bf q$, \emph{e.g.},  
the spectral density $S_V( {\bf q}; \Delta E)$ is proportional to the 
spatial Fourier transform of the potential, as to be expected from the Born 
approximation for the scattering process ${\bf p} \to {\bf p}' =
{\bf p} + {\bf q}$. The transport equation thus combines in a 
self-consistent way ballistic motion and scattering processes.

\subsection{Example: magnetic perturbation}

For illustration purposes, we show in figure~5 the 
magnetic near field spectrum at a distance $z = 1\,\mu$m
above a flat metallic surface. This spectrum is proportional to 
$S_V( {\bf q}; \Delta E)$ in~(\ref{eq:transport-eq}) for a planar
magnetic waveguide.
\begin{figure}
\begin{center}
\resizebox{7.0cm}{!} {\rotatebox{0}{
   \includegraphics{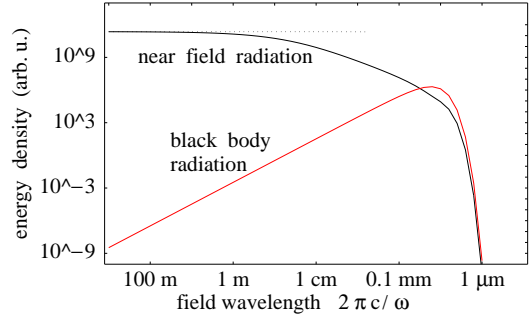}
   }}
\end{center}
\caption{{\bf Fig.\ 5.}
Magnetic field spectrum vs.\ frequency at distance $1\,\mu$m above
at copper surface at $300\,$K. The frequency $\omega$ is expressed
via the electromagnetic wave length $\lambda = 2\pi c / \omega$.
Note: a kinetic energy of $10\,\mu$K corresponds to a wave length $14\,$km.
}
\label{fig:5}
\end{figure}
One observes that for typical kinetic energies of cold atoms, the
magnetic spectrum is essentially flat. In the following, we shall
hence approximate the perturbing field by a white noise.

In figure~6,
\begin{figure}
\begin{center}
\resizebox{10.5cm}{!} {\rotatebox{0}{
   \includegraphics{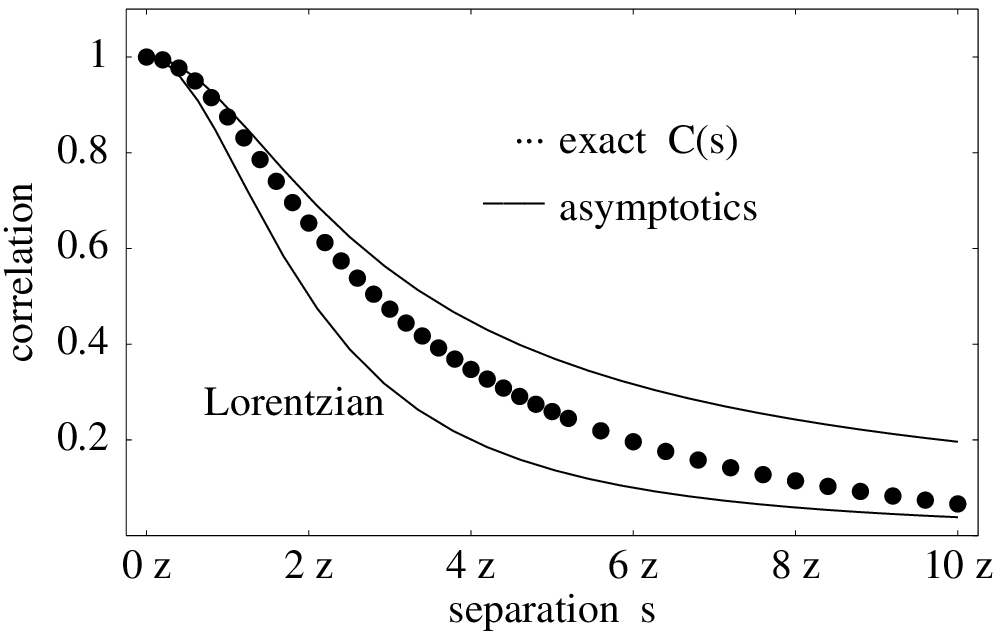}
   \hspace*{10mm}
   \includegraphics{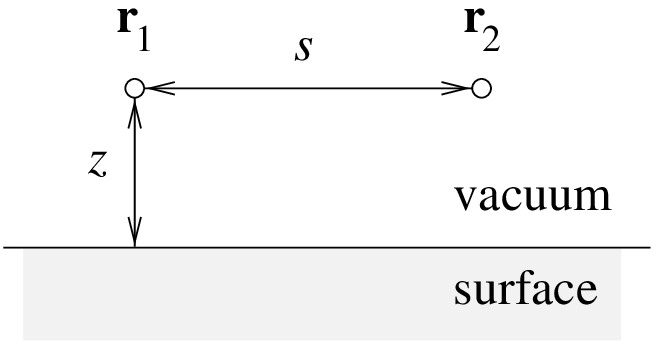}
   }}
\end{center}
\caption{{\bf Fig.\ 6.}
Spatial (normalised) correlation function of the thermal magnetic near 
field above a metallic surface at frequency $\omega/2\pi = 30\,$MHz. 
The separation $s$ gives the distance between two observations points 
at the same height $z$ above the surface.   
The other parameters are identical to the previous figure~5. 
Dots: exact evaluation, solid lines: asymptotic expansions in the
short-distance regime. For even lower frequencies, the
correlation function is essentially unchanged. 
}
\label{fig:6}
\end{figure}
we show the normalised spatial correlation function of the thermal 
magnetic field above a metallic surface.
One observes that in a planar wave guide above the surface, 
the field's spatial correlation length is of the order of the height $z$.
The correlations decay algebraically with the lateral separation $s$
(measured in the waveguide plane parallel to the surface).

\subsection{Analytic solution of the transport equation}

For a broad band spectrum of the perturbation, we may neglect the
dependence of $S_V( {\bf q}, \Delta E )$ on $\Delta E$. The transport
equation~(\ref{eq:transport-eq}) then simplifies because the integration 
over the scattered momentum ${\bf p}'$ is not restricted by energy
conservation. Taking the Fourier transform with respect to both
variables ${\bf r}$ and ${\bf p}$ (with conjugate variables ${\bf k}$
and ${\bf s}$), it is simple to derive the following solution
\begin{eqnarray}
&& \tilde{W}( {\bf k}, {\bf s}; t ) =
\tilde{W}_0( {\bf k}, {\bf s} - \hbar{\bf k}t/m ) 
\,
{\rm e}^{ - {\rm i} {\bf F}_{\rm ext} \cdot {\bf s} t / \hbar }
\times {}
\nonumber\\
&& {} \times  
\exp{\left[
- \gamma \int_0^t\!\!
\left(
1 - C( {\bf s} - \hbar{\bf k} t' /m )
\right)\!{\rm d}t'
\right]
}
\label{eq:analytic-solution}
\end{eqnarray}
Here, $\tilde{W}_0( {\bf k}, {\bf s})$ is the double Fourier transform
of the Wigner function at initial time $t = 0$, 
and $\gamma$ and the normalised spatial correlation function $C({\bf s})$ 
are related to the correlation function of the perturbation by
\be
\gamma C( {\bf s} ) =
\frac{ 1 }{ \hbar^2 }
\int_{-\infty}^{+\infty}\!d\tau\,
\langle V( {\bf s}, \tau ) \, V( {\bf 0}, 0 ) \rangle
,\quad
C( {\bf 0} ) = 1. 
\label{eq:13a}
\ee
We also note that $\gamma$ is the scattering rate
$\gamma( {\bf p} \to {\bf p}')$ for `forward scattering' processes 
where the final momentum ${\bf p}'$ approaches the initial ${\bf p}$.
For a magnetic wave guide above a metallic surface, the rate $\gamma$
is essentially of the same order of magnitude as the spin flip rate
shown in figure~4.

From the analytic solution~(\ref{eq:analytic-solution}), it is easily
checked that in the absence of the perturbation , 
the spatial width $\delta r^2(t)$ of a cloud increases ballistically
according to $\delta r^2(t) = \delta p^2(0) \, t^2 / m^2$ where
$\delta p(0)$ is the initial width of the cloud in momentum space
(this latter width remains constant in this case, of course).

\subsection{Discussion}

\paragraph{Spatial decoherence.}

More interesting information may be obtained for a nonzero scattering 
rate $\gamma$. Note that the spatially averaged atomic coherence function
is given by
\be
{\Gamma}( {\bf s} )
= \int\!{\rm d}^D{r} \,
\rho( {\bf r}; {\bf s} )
=
\tilde{W}( {\bf k} = {\bf 0}, {\bf s} )
\ee
The solution~(\ref{eq:analytic-solution}) therefore implies that the
spatial coherence decays exponentially with time:
\be
{\Gamma}( {\bf s}; t ) = {\Gamma}_0( {\bf s} )
\exp{\Bigl[ 
- 
\gamma t (1 - C( {\bf s} ) ) 
- {\rm i} {\bf F} \cdot {\bf s}\, t / \hbar
\Bigr]}
\label{eq:14}
\ee
The decoherence rate depends on the spatial separation between the points
where the atomic wave function is probed, and is given by 
$\gamma({\bf s} ) = \gamma (1 - C( {\bf s} ) )$. It hence saturates
to the value $\gamma$ at large separations and decreases to zero
for ${\bf s} \to {\bf 0}$. The decay of the coherence 
function~(\ref{eq:14}) is illustrated in figure~7.
\begin{figure}
\begin{center}
\resizebox{7.0cm}{!} {\rotatebox{0}{
   \includegraphics{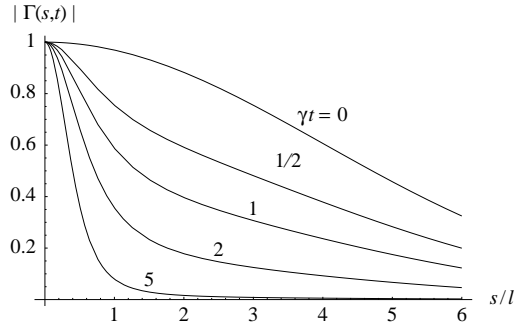}
   }}
\end{center}
\caption{{\bf Fig.\ 7.}
Illustration of spatial decoherence in an atomic wave guide. 
The spatially averaged coherence function $\Gamma({\bf s}, t)$
is plotted vs.\ the separation ${\bf s}$ for a few times $t$.
Space is scaled to the field correlation length $\ell$ and time
to the scattering time $1/\gamma$. A Lorentzian correlation
function for the perturbation is assumed.}
\label{fig:decoh}
\end{figure}
One observes that at time scales $t \ge 1/\gamma$, the spatial 
coherence is reduced to a coherence length $\xi_{\rm coh} \sim \ell$.
After a few collisions with the fluctuating magnetic field, the 
long-scale coherence of the atomic wave function is thus lost and
persists only over scales smaller than the field's correlation length
(where different points of the wave function `see' essentially the
same fluctuations). 
For larger times $t \gg 1/\gamma$, decoherence proceeds at a smaller
rate that is related to momentum diffusion, as we shall see now.

\paragraph{Momentum diffusion at long times.}

The behaviour of the atomic momentum distribution at long times
may be extracted from an expansion of the analytic 
solution~(\ref{eq:14}) for small values of ${\bf s}$. Assuming
a quadratic dependence of the field's correlation function,
$C({\bf s}) \approx 1 - s^2/\ell^2$, as one would expect for
Lorentzian correlations, we find that the atomic momentum 
distribution is gaussian at long times; it is centered at ${\bf p}_0 
+ {\bf F}_{\rm ext} t$ due to the external force, and its width
increases according to a diffusion process in momentum space
\be
\delta p^2(t) \approx \delta p^2(0) + \frac{ \hbar^2 \gamma t }{ \ell^2 }
\label{eq:p-diffusion}
\ee
This was to be expected: the atoms perform a random walk in momentum
space, exchanging a momentum of order $\hbar/\ell$ per scattering
time $1/\gamma$. The momentum diffusion coefficient $D_p = \hbar^2 \gamma
/ \ell^2$ that may be read off from~(\ref{eq:p-diffusion}) is consistent
with this intuitive interpretation. Physically speaking, the atomic
cloud is `heated up' due to the scattering from the fluctuating potential.
We note that the rate of change of the atomic kinetic energy in the
wave guide plane is the same as the one for the tightly bound motion 
perpendicular to the metallic surface (see \cite{Henkel99c} for a 
calculation of this rate).

Translating the width of the momentum distribution into a spatial coherence 
length, we find a power-law decay at long times, $\xi_{\rm coh} = 
\ell / \sqrt{ \gamma t }$. 
Finally, a similar calculation yields the width of the atomic cloud 
in position space: it increases `super-ballistically' at long times, 
$\delta r^2(t) \propto t^3$, as a consequence of heating.

\section{Conclusion}

Particles in small traps close to macroscopic bodies are subject
to fluctuating near fields that show noise spectra orders of magnitude
above the blackbody level. This is because the geometric distances involved
are typically much smaller than the electromagnetic wave lengths
associated with the relevant frequencies. As a consequence, the ground
state of the vibrational motion of trapped ions is unstable, and coherences
between different oscillator levels decay. We have developed a theoretical
framework to compute the corresponding heating and decay rates. 
As a second consequence, quasi-free matter waves in a linear or planar
wave guide in the vicinity of macroscopic bodies (substrates or wires)
are scattered and lose their spatial coherence. We have identified
the relevant time scale and obtained analytic estimates for the behaviour
of the atomic coherence function at large spatial and temporal scales.
These estimates should be useful, we hope, to design integrated atom
optical circuits with controlled decoherence. 

Questions that could be addressed in the future pertain to detailed
theories for trapped ion heating, as well as to transport processes for
condensed atomic samples. Finally, the inclusion of interference effects 
in multiple scattering might provide a link to study weak and/or 
strong localisation of matter waves in wave guides.

\medskip
\noindent {\bf Acknowledgments.} C.H. would like to thank R.\
Carminati and K.\ Joulain of the group of J.-J.\ Greffet at
Ecole Centrale (Paris) for useful discussions. Furthermore, 
D.\ Leibfried, E. Peik, F.\ Schmidt-Kaler, J. Schmiedmayer,
and J.\ von Zanthier communicated precious hints on experimental issues.


\bibliography{/Net/Users/carstenh/Biblio/Database/journals,/Net/Users/carstenh/Biblio/Database/biblioac,/Net/Users/carstenh/Biblio/Database/bibliodh,/Net/Users/carstenh/Biblio/Database/biblioio,/Net/Users/carstenh/Biblio/Database/bibliopz}
\bibliographystyle{/Net/Users/carstenh/Biblio/Database/bst/actaslov}



\end{document}